\documentclass[prl,aps,twocolumn,preprintnumbers,showpacs,superscriptaddress,epsfig]{revtex4-1}
\usepackage{amsmath,amsfonts,amssymb,amsthm,graphics,graphicx,epsfig,times,bbm}
\usepackage[colorlinks=true,citecolor=blue,linkcolor=magenta]{hyperref}
\usepackage[usenames]{color}
\usepackage{graphicx}
\usepackage{subfigure}
\usepackage{amsmath}
\usepackage{epsfig,times}
\usepackage{dcolumn}
\usepackage{bm}
\usepackage{color}
\usepackage{verbatim}
\usepackage{epstopdf}
\usepackage{amssymb}
\usepackage{amstext}
\usepackage{latexsym}
\usepackage{hyperref}
\usepackage{amsfonts}
\usepackage{psfrag}
\usepackage{verbatim}
\usepackage{xcolor}

\usepackage{mathptmx} 

\usepackage[colorlinks=true,citecolor=blue,linkcolor=magenta]{hyperref}

\def\V{{\cal V}}
\def\A{{\cal A}}
\def\ln{{\rm ln}}

\def\t{\hat{\rm\bf t}}
\def\n{\hat{\rm\bf n}}
\def\b{\hat{\rm\bf b}}

\def\U{{\rm U}}

\def\d{{\rm d}}

\def\om{\omega}

\newcommand{\beqa}{\begin{eqnarray}}
\newcommand{\eeqa}{\end{eqnarray}}

\begin{document}

\title{
Bent waveguides for matter-waves:  supersymmetric  potentials and reflectionless geometries
}

\author{Adolfo del Campo$^{1,2,*}$, Malcolm G. Boshier$^{3}$, and Avadh Saxena$^{1,2}$}

\affiliation{$^{1}$Theoretical Division,  Los Alamos National Laboratory, Los Alamos, NM 87545, USA\\
$^{2}$Center for Nonlinear Studies,  Los Alamos National Laboratory, Los Alamos, NM 87545, USA\\
$^{3}$Physics Division, Los Alamos National Laboratory, Los Alamos, NM 87545, USA\\
$^*$ Corresponding author: \texttt{delcampo@lanl.gov}}

\begin{abstract}
Non-zero curvature in a waveguide leads to the appearance of an attractive quantum  potential which crucially affects the dynamics in matter-wave circuits.
Using methods of supersymmetric quantum mechanics, pairs of bent waveguides are found whose geometry-induced potentials share the same scattering properties. As a result, reflectionless waveguides, dual to the straight waveguide, are identified. Strictly isospectral waveguides are also found by modulating the depth of the trapping potential. Numerical simulations are used to demonstrate the efficiency of these approaches in tailoring and controlling curvature-induced quantum-mechanical effects.

\end{abstract}



\maketitle

\section{Introduction}
Waveguides with non-zero curvature  are basic constituents of matter-wave circuits in atom chip technology \cite{atomchip,chips}, as well as its ion \cite{ionchip}, 
molecular \cite{molecularchip}, 
and electron \cite{electronchip} counterparts.
Their relevance is further enhanced by the development of flexible techniques to create optical waveguides for ultracold gases. 
In this context, waveguide trapping potentials  can be engineered by a variety of methods including the time-averaging painted potential technique \cite{painters}, the use of an intensity mask \cite{Anderson07,Heinzen09}, and holographic methods, in particular, digital holography \cite{GH12}.
Circular ring traps have attracted a considerable amount of attention \cite{ring1,ring2,ring3,ring4,ring5,ring6,painters}, most recently to study Josephson junction dynamics \cite{Wright2013, Ryu13}.  Other curved waveguides have also been engineered, such as a stadium-shaped potential trap \cite{stadium04,Heller06}.

The propagation of matter-waves in bent waveguides generally differs from that in straight waveguides due to the appearance of a purely attractive local quantum potential of geometrical origin \cite{SRS77,daCosta81,daCosta82}. Under tight-transverse confinment, the magnitude of this curvature-induced potential (CIP)   is proportional to the square of the curvature of the waveguide, and affects both the single-particle and many-body physics of the confined matter-waves  \cite{EB89,GJ92,CB96,Clark98,EV99,LP01,Schwartz06}. As a result, the scattering properties of a curved  tight waveguide are modified,  e.g., by the  appearance of bound states \cite{EB89,GJ92}. Advances in the design of bent waveguides, in which curvature-induced effects are tailored and suppressed, are required for the miniaturization of matter-wave circuits. It is to this problem that we turn our attention.
%

\section{Results}
In this manuscript we design bent waveguides for matter-wave circuits free from spurious quantum mechanical effects associated with CIPs. 
Three novel ideas are presented: 
(i) Exploiting the interplay of geometry and supersymmetry in quantum mechanics, we relate pairs of waveguides whose 
 CIPs are isospectral  and share the same scattering properties. 
(ii) We then identify waveguides which are reflectionless for coherent  matter-waves at all energies.
(iii) Furthermore, we show that by tailoring the depth of the waveguide trap, it is possible to cancel the CIP, rendering the dynamics of the guided matter-waves equivalent to that in straight waveguides.

Let us consider the dynamics of matter-waves
confined in a tight waveguide whose axis follows the curve $\gamma$, parametrized as a function of the arc length $q_1$ by the vector 
${\bf r}={\bf r}(q_1)$, with tangent ${\bf t}(q_1)$. 
We start by recalling the fundamental theorem of curves which asserts that a curve is completely determined, up to its position in space, by its curvature $\kappa$ and torsion $\tau$ \cite{Struik88}.
Indeed, the expressions $\kappa=\kappa(q_1)$ and $\tau=\tau(q_1)$ constitute the natural intrinsic equations of a curve. 
A parametrization of the curve can be obtained by integration of the Frenet-Serret equations
\beqa
d_{q_1}
\begin{pmatrix}
\t \\
\n  \\
\b \\
 \end{pmatrix}
=
\begin{pmatrix}
0 & \kappa &  0\\
-\kappa & 0 & \tau  \\
0 & -\tau & 0 \\
 \end{pmatrix}
\begin{pmatrix}
\t \\
\n  \\
\b \\
 \end{pmatrix},
\eeqa
where $\n=(d\t/dq_1)/\kappa$ (provided $d\t/dq_1\neq 0$) and $\b=\t\times\n$ are the  principal normal and binormal unit vectors, and the curvature and torsion at the point of arc length $q_1$ are defined as $\kappa(q_1)=|\d{\bf t}(q_1)/dq_1|$ and $\tau(q_1)=-\n\cdot d\b/dq_1$, respectively.
Let $(q_2,q_3)$ be the transverse local coordinates
and consider a transverse confining potential 
 $\U_{\lambda}(q_2,q_3)$ such that in the limit of tight confinement $\lambda\rightarrow\infty$ the particle is bounded to $\gamma$.
Under dimensional reduction, the purely-attractive CIP emerges \cite{daCosta81,daCosta82,EB89,GJ92,CB96} 
\beqa
\label{daCostapot}
\V(q_1)=-\frac{\hbar^2}{8m}\kappa(q_1)^2.
\eeqa
This result is independent of $\U_{\lambda}$ and holds in particular under an isotropic transverse harmonic confinement $\U_{\om_{\perp}}=m\om_{\perp}^2(q_2^2+q_3^2)/2$ with ground state width $\sigma_0=[\hbar/(m\om_{\perp})]^{1/2}$ \cite{LP01,Schwartz06}. The conditions for the  dimensional reduction to be valid explicitly read
\beqa
\label{cc}
\kappa \sigma_0\ll 1,\qquad |\kappa'|\sigma_0\ll|\kappa|,\qquad |\kappa''|\sigma_0\ll\kappa^2,
\eeqa
where primes denote derivatives with respect to $q_1$.

We next pose the problem of identifying pairs of waveguides with the same scattering properties, and engineering a waveguide which minimizes the effect of the CIP.
Generally, direct integration of the Frenet-Serret equations is not possible. However, the quantum mechanical behavior of matter waves bounded to  isometric curves 
with different torsion but the same curvature remains the same \cite{daCosta81,daCosta82} because $\V$ is independent of $\tau$. In addition, 
matter-wave circuits in atom chips and optical realizations of waveguides are often associated with curves $\gamma$ on a plane, for which $\tau=0$. As a result, 
we  focus on planar curves, 
given by the parametrization ${\bf r}(q_1)=(x(q_1),y(q_1))$.  
We shall return to the case of $\tau\neq0$ whenever the curve $\gamma$ exhibits multiple points in which the waveguide self-intersects.
The CIP depends only on $\kappa^2$ and remains invariant under 
the mapping  
$\kappa\rightarrow {\rm sgn}(g(q_1))\kappa$ with an arbitrary real function $g(x)$, a symmetry which we shall exploit to engineer the guiding potential.
Provided $\tau=0$, and $\kappa(q_1)\neq 0$ for all $q_1$ it is always possible to integrate the Frenet-Serret equations, 
and to find the natural representation of the curve in terms of the arc length
\beqa
\label{IFSE}
\nonumber x(q_1)&=&x_0+\int_{q_1^0}^{q_1}\cos\left(\int_{q_1^0}^{\bar{s}}\kappa(s)ds\right)d\bar{s},\\
\\
\nonumber y(q_1)&=&y_0+\int_{q_1^0}^{q_1}\sin\left(\int_{q_1^0}^{\bar{s}}\kappa(s)ds\right)d\bar{s}.
\eeqa
In what follows we shall take $(x_0,y_0)=(0,0)$ without loss of generality, and extend Eq. (\ref{IFSE}) to $q_1<0$ to sample  the full range of $\V$.

\subsection{Supersymmetric partner waveguides}
In the Witten model of supersymmetric quantum mechanics (SUSY QM) \cite{Witten81,CKS95}, a pair of SUSY partner Hamiltonians is considered with the factorization,
\beqa
H_- :=\A^{\dag}\A\geq 0,\qquad  H_+:=\A\A^{\dag}\geq 0,
\eeqa
where the annihilation and creation operators are defined by
$\A:=\frac{ip}{\sqrt{2m}}+\Phi(q_1)$ and $\A^{\dag}:=-\frac{ip}{\sqrt{2m}}+\Phi(q_1)$, in terms of the superpotential $\Phi(q_1)$. 
The SUSY partner Hamiltonians can be explicitly written as
$H_{\pm}=-\frac{\hbar^2}{2m}\partial_{q_1}^2+V_\pm$, where $V_\pm(q_1):=\Phi^2(q_1)\pm\frac{\hbar}{\sqrt{2m}}\Phi'(q_1)$ are partner potentials.
We consider the case in which the SUSY partner potentials are both induced by curvature, i.e. $V_\pm=\V_\pm=-\frac{\hbar^2}{8m}\kappa_\pm^2$. It follows that SUSY partner Hamiltonians are associated with curves whose curvatures $\kappa_{\pm}$ are related by 
\beqa
\label{SUSYkappa}
\kappa_{+}^2(q_1)&=&\kappa_{-}^2(q_1)-\frac{8\sqrt{2m}}{\hbar}\Phi'(q_1),\\
&=&\kappa_{-}^2(q_1)-\frac{8m}{\hbar^2}[\A,\A^{\dag}],
\eeqa
where we have used the relation between the derivative of the superpotential and the commutator of the creation and annihilation operators in the second line \cite{DKS88}.
The relation between $\kappa_\pm$, can be further developed by making reference to the ground state $\psi_0$ of $H_-$ satisfying $H_-\psi_0=0$, in terms of which 
the superpotential reads $\Phi(q_1) =-\frac{\hbar}{\sqrt{2m}}\frac{\partial_{q_1}\psi_0}{\psi_0}$. Using this expression in  (\ref{SUSYkappa}), it follows that
\beqa
\label{SUSYkappa2}
\kappa_{+}^2(q_1)=\kappa_{-}^2(q_1)+8\bigg[\frac{\partial_{q_1}^2\psi_0}{\psi_0}-\left(\frac{\partial_{q_1}\psi_0}{\psi_0}\right)^2\bigg].
\eeqa
According to the fundamental theorem of curves, the shape and length of a planar curve is completely determined by its (single-valued and continuous) curvature. 
It follows that the SUSY partner potentials $\V_\pm$  are associated with the family of curves  $\{\gamma_\pm\}$ 
(with curvatures $\kappa_\pm$ satisfying $\kappa_\pm^2=-\frac{2m}{\hbar}\Phi^2\mp\frac{\sqrt{2m}}{\hbar}\Phi'$), 
that we shall refer to as SUSY partner curves. This set can  be extended to include  curves associated with a family of  shape-invariant potentials \cite{Gendenshtein83}, as discussed in Methods,  
or using  higher-order SUSY QM \cite{CKS95}.

\begin{figure}
\centering{\includegraphics[width=0.7\linewidth]{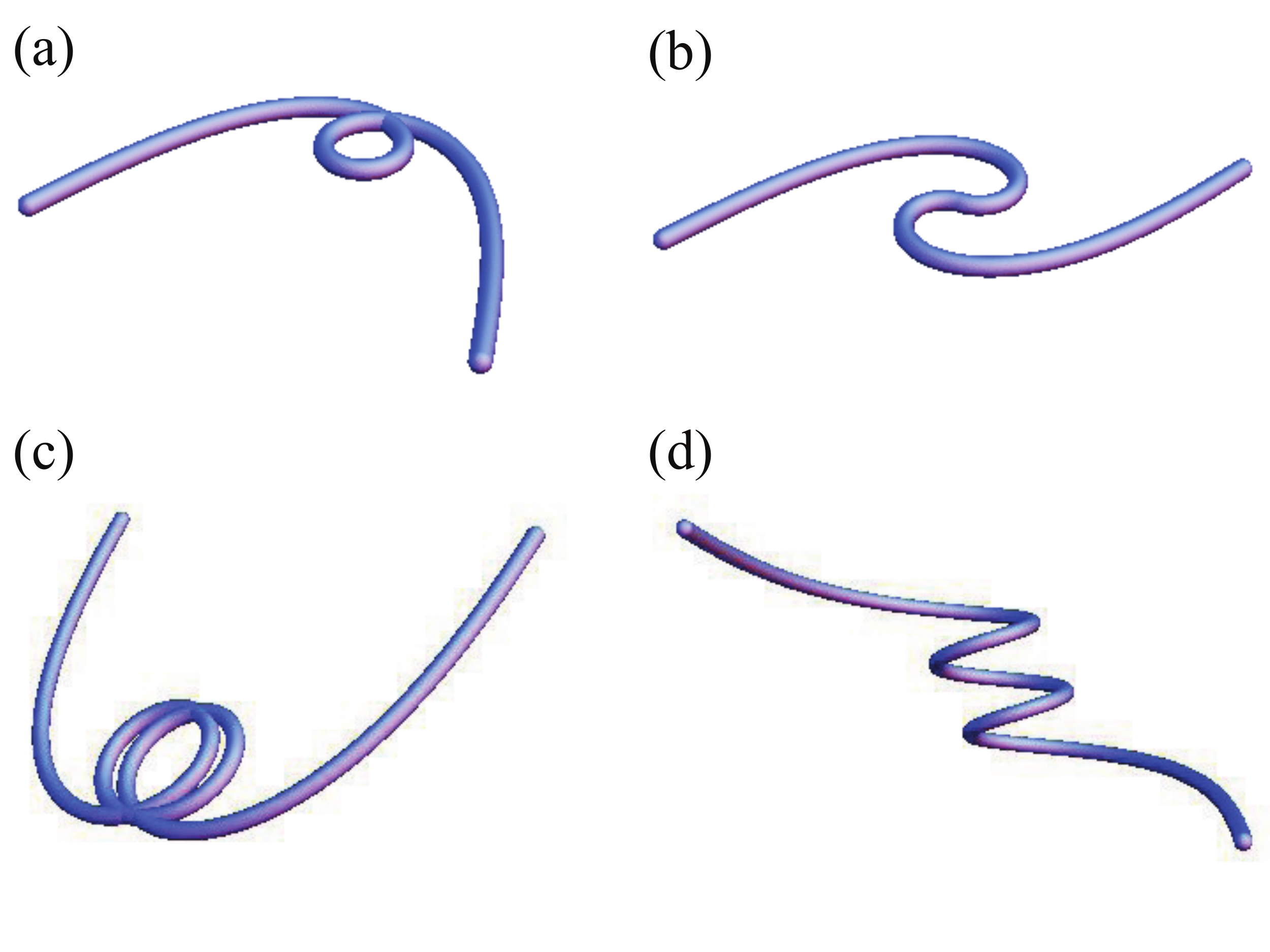}}
\caption{
{\bf  Supersymmetric reflectionless waveguides.}
(a) Waveguide whose CIP is the reflectionless P\"oschl-Teller potential with $\nu=1$. Its SUSY partner curve is the straight line $\gamma_-$ with zero curvature.  
(b)  The multiple point  can be removed to obtain a simple waveguide using the curvature mapping $\kappa(q_1)\rightarrow {\rm sgn}(q_1)\kappa(q_1)$ under which the CIP remains invariant. 
(c) For higher curvature values, as in the waveguide associated with the $n=2$ Sukumar reflectionless potential shown here,  integration of the Frenet-Serret equations leads to curves with several multiple points. 
(d) Such CIPs can be engineered  in a non-planar waveguide,  with non-zero torsion $\tau$.
}
\label{ReflectionlessCurves}
\end{figure}

\begin{figure*}
\centering
\centering{\includegraphics[width=0.95\linewidth]{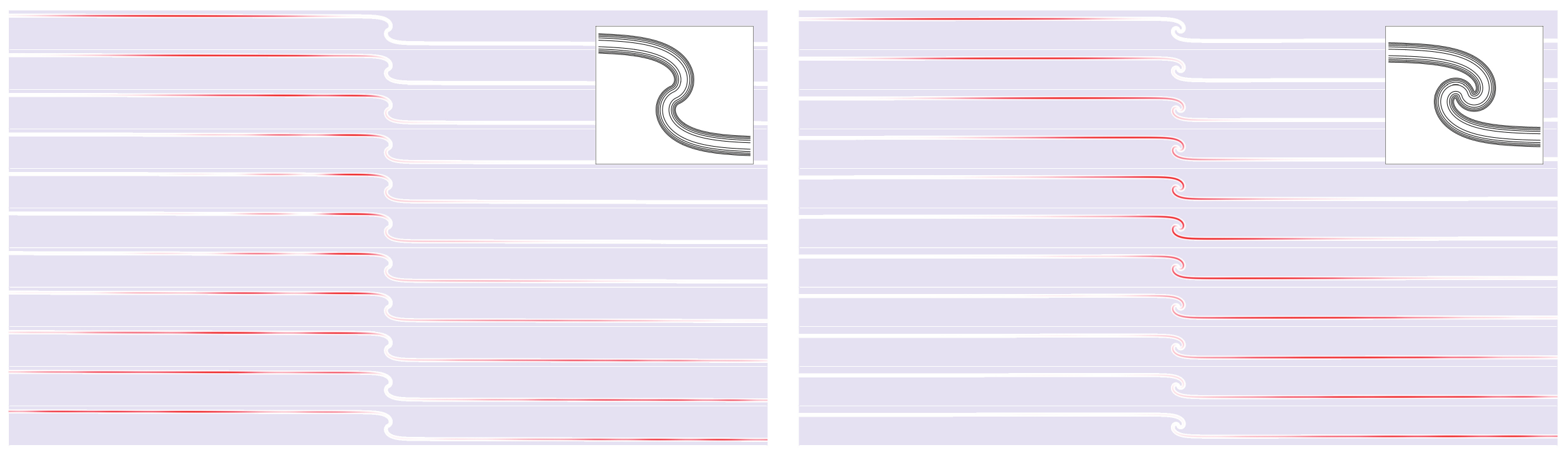}}
\caption{
{\bf  Scattering dynamics in bent waveguides.}
Sequence of snapshots of the time-evolution of the density profile of a wavepacket  along a planar bent waveguide with the curvature  (\ref{MPTkappa}) and $\nu=1/2$ (left), and $\nu=1$ (right), as in Fig.\,\ref{ReflectionlessCurves}(b).  
Generally, the wavepacket is split by the CIP into a transmitted and a reflected component. 
Despite the high degree of bending shown in the inset, whenever $\nu$ is an integer, the waveguide becomes reflectionless and exhibits unit transmission probability for all energies of the impinging matter-wave beam.  
The color coding varies from white to red as the probability density increases.The dimensions of each waveguide image are $908\sigma_0 \times 47\sigma_0$ and the time interval between successive images is $1920/\om_{\perp}$.  The initial wavepacket has FWHM = 235$\sigma_0$ and momentum $(1/32)m\om_{\perp}\sigma_0$, and  
$\alpha = 1/8$.
}
\label{ReflectionlessScattering}
\end{figure*}

Why are waveguides along these curves of interest? The main physical feature of SUSY partner curves is that they exhibit the same scattering properties, a distinguishing feature directly inherited from $\V_\pm$ \cite{CKS95,propagators}. Let  $\gamma_\pm$ be open waveguides with finite curvature as $q_1\rightarrow\pm\infty$, so that $\V_{\pm}(q_1\rightarrow\pm\infty)\rightarrow\Phi(q_1\rightarrow\pm\infty)^2=:\Phi_{\pm}^2$,  and consider the scattering states of momentum $k$ and energy $E=\hbar^2k^2/(2m)$, with reflection and transmission amplitudes $R_{\pm}(k)$ and $T_{\pm}(k)$, respectively. It follows that $R_-(k)=\frac{\Phi_-+i\hbar k/\sqrt{2m}}{\Phi_- -i\hbar k/\sqrt{2m}}R_+(k)$ and $T_-(k)=\frac{\Phi_+-i\hbar k'/\sqrt{2m}}{\Phi_- -i\hbar k/\sqrt{2m}}T_+(k)$ where $k=[2m(E-\Phi_-^2)]^{1/2}/\hbar$ and $k'=[2m(E-\Phi_+^2)]^{1/2}/\hbar$, that is, the reflection as well as the transmission probabilities are the same for SUSY partner curves.
Further, the Hamiltonians $H_\pm$ associated with SUSY partner curves are isospectral, except for the lowest energy level of $H_-$ with zero-energy, which is absent in the spectrum of $H_+$.

%
%

\subsection{Design of reflectionless curves}
 
CIPs are of attractive character and as a result can  lead to quantum reflection \cite{PSU58a,PSU58b,Henkel96,FT04}.
The dynamics of a guided matter-wave on a bent waveguide is generally  affected by the curvature.
We next illustrate the power of the SUSY partner waveguides in designing reflectionless curves.
An obvious instance where the CIP vanishes is that of an infinite straight waveguide, with $\kappa_-=0$ and superpotential $\Phi=A\tanh\alpha q_1$ with $A>0$.  
This configuration is of relevance to guided atom lasers \cite{GAL1,GAL2}, and we wish to mimic it in bent waveguides.
SUSY QM allows us to find SUSY partners which are reflectionless.   In this case, $V_+(q_1)$ is given by the modified P\"oschl-Teller potential $V_{\rm curv}(q_1)=-\frac{\hbar^2}{2m}\frac{\nu(\nu+1)}{\cosh^2(\alpha q_1)}$ \cite{CKS95}, so that the curvature of the SUSY $\gamma_+$ curve reads
\beqa
\label{MPTkappa}
\kappa_+(q_1)=2\alpha\sqrt{\nu(\nu+1)}{\rm sech}\alpha q_1, 
\eeqa
where $\nu$ is a positive integer.
Provided that the dimensional reduction is valid, the transmission probability for  a waveguide  with curvature  (\ref{MPTkappa}) and  arbitrary $\nu$,  is given by 
$|T_+|^2(k)=\frac{\mu^2}{1+\mu^2}$ with $\mu=\frac{{\rm sinh}(\pi k/\alpha)}{\sin \pi \nu}$ and $k=\sqrt{2mE/\hbar}$. Such a waveguide becomes reflectionless for integer values of $\nu$.
Different reflectionless waveguides are plotted in Fig. \ref{ReflectionlessCurves}, where it is shown that  the number of multiple points increases with the magnitude of the curvature.
A simple waveguide without junctions can then be engineered by exploiting the invariance of the CIP with respect to changes in the sign of the curvature, or by considering a nonzero torsion $\tau\neq0$, whose realization might be achieved using an extended version of the painted potential technique \cite{painters}. 
We emphasize that there are infinitely many instances of reflectionless waveguides.
Let us illustrate this by adapting the algorithm developed by Sukumar \cite{Sukumar86} to CIPs.  One can construct reflectionless waveguides supporting $n$ bound states.
Let us fix the bound state energies to be $E_n=-\frac{\hbar^2}{2m}\eta_n^2$ with $\eta_n^2>\eta_{n-1}^2>\dots\eta_1^2$.
The symmetric reflectioneless curves are described by the equation 
\beqa
\kappa_n^2(q_1)=8\partial_{q_1}^2\ln \det D_n,
\eeqa
 where $[D_n]_{ij}=\frac{1}{2}\eta_j^{i-1}[\exp(\eta_jq_1)+(-1)^{i+j}\exp(-\eta_jq_1)]$.
For a single bound state, one obtains $\kappa^2=8\eta_1^2{\rm sech}\eta_1q_1$, closely related to the SUSY curves associated with (\ref{MPTkappa}). 
Figure  \ref{ReflectionlessCurves} (lower panels)  shows the  reflectionless curves corresponding to a Sukumar potential supporting two bound states with $\eta_1=1$ and $\eta_2=3/2$, where multiple points in  \ref{ReflectionlessCurves}(c) are avoided by a non-zero torsion ($\tau=20q_0^{-1}$) in \ref{ReflectionlessCurves}(d).  
The axis of the associated waveguide follows the curve $(x(s),y(s),\tau s)$ with (squared) curvature $\kappa^2(s)=-8m\V(s)/\hbar^2$, torsion $\tau$ and arc length $q_1=\sqrt{1+\tau^2}s$. At variance with (\ref{MPTkappa}), the relative angle between the asymptotes can be tuned by adjusting the value of $\eta_2$ relative to $\eta_1$, which will allow for the engineering of reflectionless bends through a range of desired angles. 
Further examples of reflectionless waveguides can be found by using the infinite family of reflectionless potentials discussed by Shabat \cite{Shabat92} and Spiridonov \cite{Spiridonov92}. 
The reflectionless character of the SUSY waveguides becomes apparent in the dynamics of guided matter waves. Figure \ref{ReflectionlessScattering} shows an elongated Gaussian beam being guided in a bent waveguide with curvature given by (\ref{MPTkappa}).  $\gamma_+$ is asymptotically flat for $q_1\rightarrow\pm\infty$. 
For a general non-integer value of $\nu$, the traveling beam is substantially reflected off the bent region. For integer $\nu$ there exists a delocalized critical bound-state with zero energy 
and the waveguide becomes reflectionless for all scattering energies. However, the degree of bending increases with $\nu$. As a result, reflectionless waveguides provide a remarkable counterexample to the common expectation that the reflection probability increases with the degree of bending of the waveguide.
In addition, the numerical simulations correspond to the propagation in a waveguide with finite transverse width, for which the explicit form of the curvature induced potential \cite{SRS77,EB89,Schwartz06} is more complex than that in Eq. (\ref{daCostapot}) used to design the reflectionless SUSY waveguide, and where excitations of the transverse waveguide modes are possible. The fact that despite the  finite transverse  width the waveguide remains reflectionless illustrates the robustness of its design against imperfections. We also note that the reflectivity of the P\"oschl-Teller potential changes only gradually as $\nu$ departs from an integer value.


\begin{figure}
[t]
\centering{\includegraphics[width=0.7\linewidth]{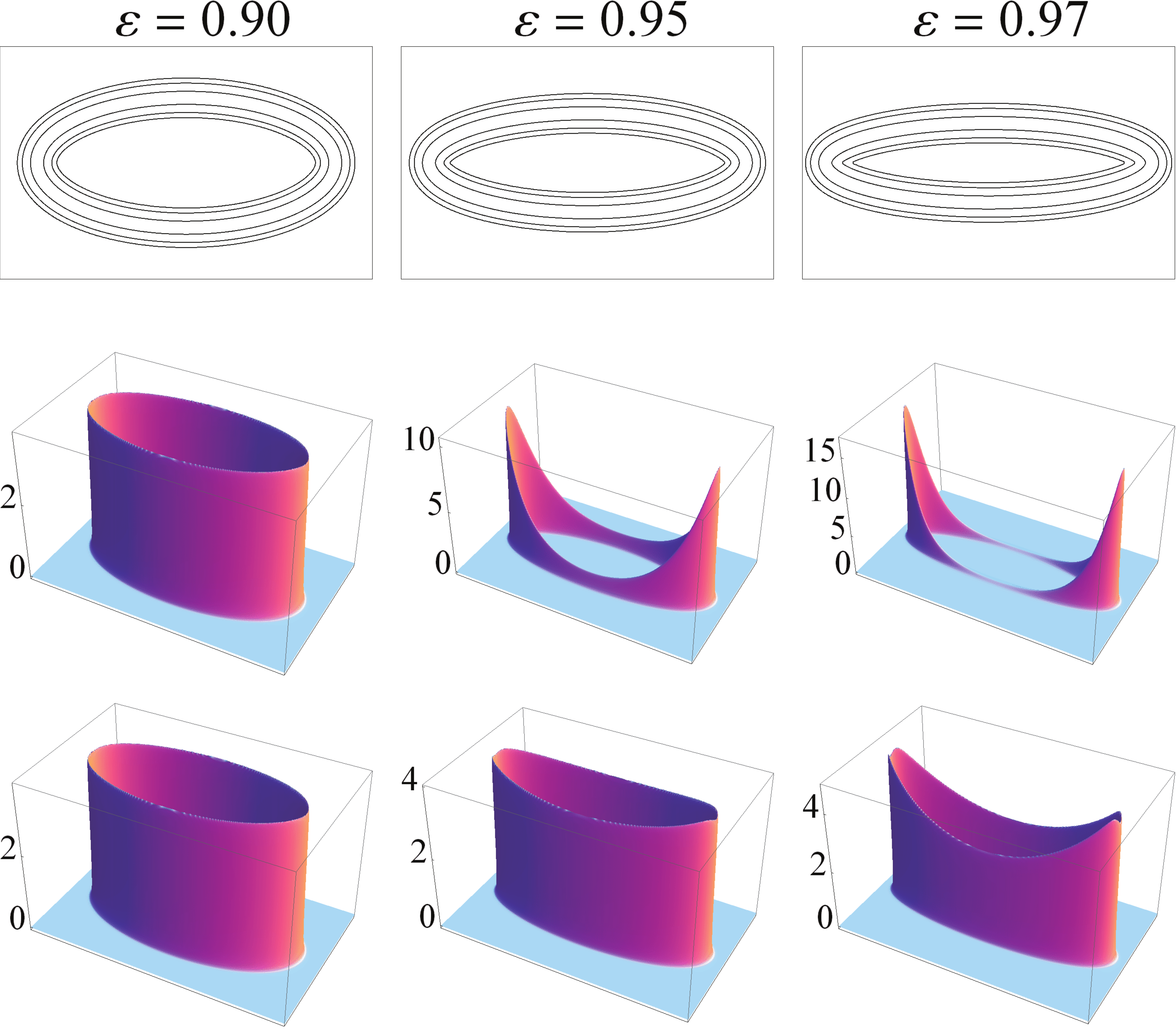}}
\caption{
{\bf  Canceling out the curvature-induced potential.}
Elliptical waveguide potentials of increasing eccentricity (top) and corresponding ground state densities (middle).  Bottom row shows ground state densities when the CIP is compensated by modulating the depth of the trap.  The dimensionless density profile $n(q_1)\sigma_0$ is scaled up by a factor $10^3$,  the perimeter of the ellipse is $L=150\sigma_0$ and the plotted area is $80\sigma_0\times 50\sigma_0$.
}
\label{noGIP}
\end{figure}

\subsection{Canceling out the geometry-induced potential}
A variety of experimental techniques to design matter-wave circuits, such as the painted potential technique, 
offers an alternative way to control the design of $\gamma$: the modulation of the potential depth of the waveguide. 
Consider two arbitrary isometric waveguides $\gamma$ and $\widetilde{\gamma}$ (both either open or closed, and without multiple points), 
with CIPs $\V_{\gamma}(q_1)$ and $\V_{\widetilde{\gamma}}(q_1)$, respectively. 
Under  the consistency conditions (\ref{cc}), it is then possible to make $\widetilde{\gamma}$ isospectral to $\gamma$ by modulating the depth of the waveguide potential, i.e.,  
by creating a potential barrier of the form $U(q_1)=-[\V_{\widetilde{\gamma}}(q_1)-\V_{\gamma}(q_1)]$. 
A natural case is that in which  $\V_{\gamma}(q_1)$ either vanishes or is an irrelevant constant energy shift. 
$U(q_1)$ is then the potential required to flatten out the  depth of the global potential of $\widetilde{\gamma}$.  
In addition, the acceleration of the guided matter-waves towards the region of high-curvature is prevented.
To explore in detail this possibility, we consider 
an elliptical trap \cite{SRS77,SK01,Schwartz06}, associated with the curve ${\bf r}(u)=(a\cos u,b\sin u)$,  with $a\geq b>0$, and circumference $L$. 
The CIP in an elliptical trap reads
\beqa
\V(u)=-\frac{\hbar^2}{8m}\frac{a^2b^2}{(b^2\cos^2u+a^2\sin^2u)^3}.
\eeqa
The eccentricity of an ellipse is defined by $\epsilon = [1-\left(b/a\right)^2]^{\frac 1 2}\in [0,1]$ and
can be used to quantify the deformation from a circle (for which $a=b$, $\epsilon=0$). 
For a ring of radius $a=b$ ($\gamma$, with $\epsilon=0$), the curvature is $\kappa(q_1)=1/a$ and the CIP becomes constant, and the ground state density profile is uniform along the arc length $q_1$. For $\widetilde{\gamma}$ with $\epsilon>0$, the CIP comes into play and creates two attractive double wells, centered around the points with higher curvature $q_1=\{0,L/2\}$ ($b<a$) and with the minimum value $-\frac{\hbar^2}{8m}\frac{a^2}{b^4}$. 
The extent to which geometry-induced effects can be cancelled out  by painting a barrier $U(q_1)=-\V_{\widetilde{\gamma}}(q_1)$ is illustrated in Fig.  \ref{noGIP}. 
Such cancellation is effective as long as the consistency conditions for the dimensional reduction hold, which ceases to be the case as $\epsilon$ is increased while the transverse width $\sigma_0$ remains fixed.
  
\begin{figure}
[t]
\centering{
\includegraphics[angle=0,width=1.1\linewidth]{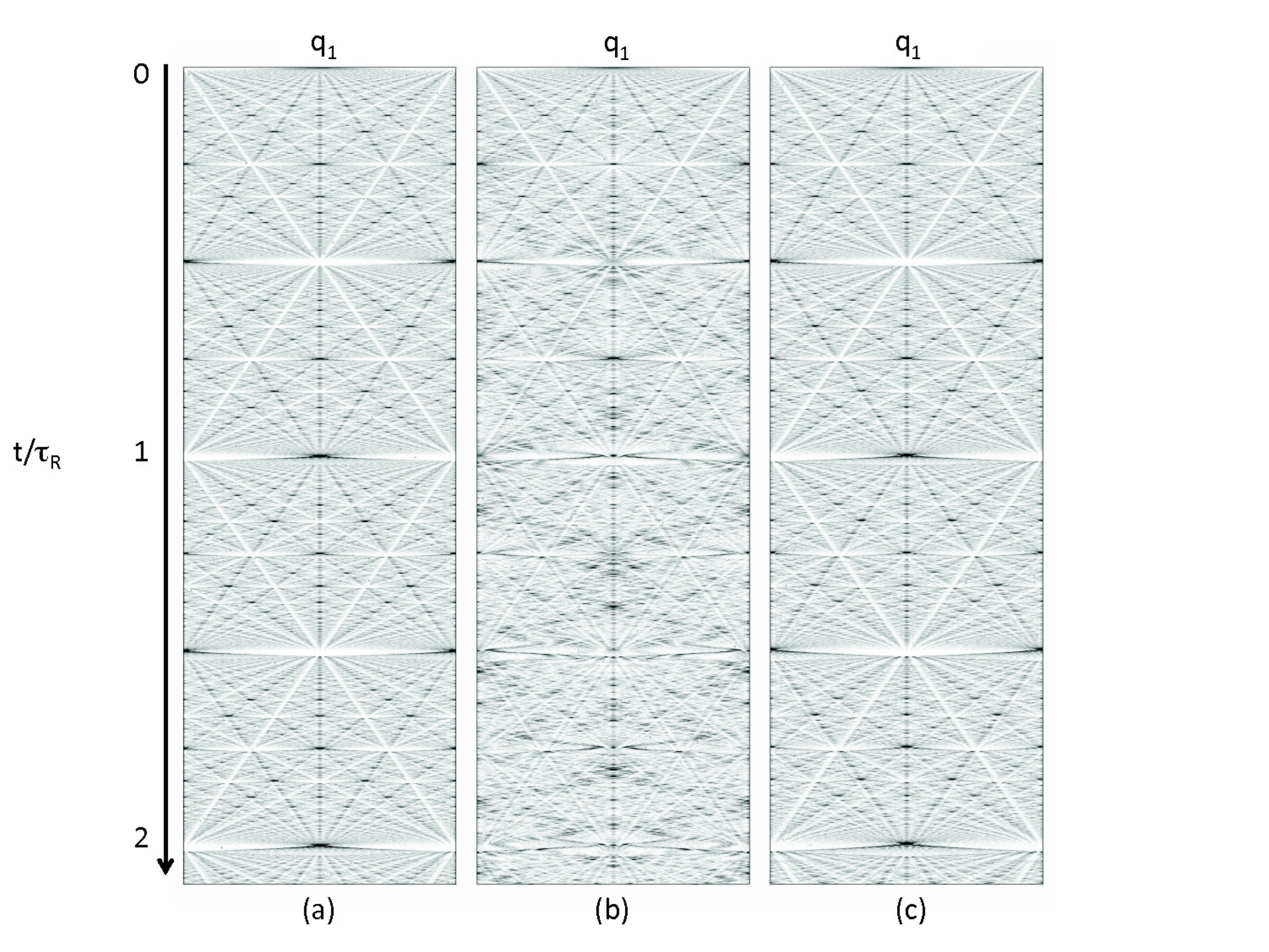}
}
\caption{
{\bf  Curvature-induced suppression of temporal Talbot oscillations.}
Time evolution of the density profile $n(q_1,t)=\int dq_{\perp}n(q_1,q_\perp,t)$ 
of an initially tightly-localized wavepacket released in a two-dimensional elliptical waveguide.
(a) For a ring trap ($\epsilon=0$) $n(q_1,t)$ exhibits Talbot oscillations as  a result of the  quadratic  dispersion relation (left).  
Two Talbot oscillations are displayed.
(b) Whenever $\epsilon>0$, the CIP lifts the degeneracies in the spectrum and suppresses Talbot oscillations ($\epsilon=0.9$). 
(c) The CIP can be cancelled out by modulating the depth of the trap ($\epsilon=0.9$).  
 $L=150\sigma_0$  in all cases and hence the revival time $\tau_R = m L^{2}/(\pi \hbar)$ is constant for different values of $\epsilon$.
}
\label{carpet}
\end{figure}

The ground state density profile is a fairly robust quantity, but we note that this compensation is efficient as well for dynamical processes involving all spectral  properties of the waveguide.
Consider the  time evolution of the density profile of an initially localized wavepacket released in  the elliptical trap, displayed in Fig. \ref{carpet}.
For a ring trap, where $\epsilon=0$, the evolution of the density profile $n(q_1,t)$  weaves a highly structured interference pattern with ``scars'' in the plane $(q_1,t)$, known as a ``quantum carpet'' \cite{Berry96}. Such quantum carpets exhibit a temporal analogue of the Talbot effect: in wave optics, the near-field diffraction pattern of a  wave incident upon a periodic grating is characterized by a spatial periodicity \cite{Talbot1836,BMS01}.  The quantum dynamics of an initially localised wavepacket which is released in a two-dimensional ring trap exhibits 
 a periodic revival of the initial state, with period  $\tau_R=mL^2/(\pi\hbar)$ (the temporal analogue of optical Talbot oscillations). The reconstruction of the density profile at $t=0$ can be traced back to the quadratic dispersion relation of the trap and the degeneracies it entails \cite{Schleich1,Schleich2}. This phenomenon has been experimentally observed in a variety of systems \cite{Chapman95,Mark11}. In a  two-dimensional elliptical waveguide, the spectrum is modified and the dispersion relation ceases to be quadratic. For $\epsilon>0$, the geometry-induced potential lifts the degeneracy in the spectrum, leading to the suppression of Talbot oscillations. 
Nonetheless, the dynamics corresponding to a ring trap can be effectively recovered in an ellipcal trap with $\epsilon>0$ 
after compensating the depth of the waveguide potential. The reapperance of Talbot oscillations in compensated elliptical waveguides signals the isospectral properties with respect to the ring trap, illustrating the suppression of curvature-induced effects. 



\section{Discussion}
The dynamics of matter waves in bent waveguides is severely distorted by the appearance of an attractive curvature-induced quantum potential.  As matter wave circuits shrink in size and atomic velocities must be reduced to maintain single mode propagation, curvature-induced potentials impose practical limitations on minimum velocities, and  methods to reduce their effects are needed.
Using methods of supersymmetric quantum mechanics, we have introduced a framework to design sets of bent waveguides which share the same scattering properties. 
As a relevant example, an infinite family of reflectionless waveguides with a controllable number of bound states has been presented. As a complementary approach, 
we have discussed the possibility of tailoring curvature-induced effects by controlling the depth of the  waveguide trapping potential.
Our discussion has been focused on the  effects of curvature on guided  matter-waves which are experimentally realizable by a  variety of techniques including atom chip technology 
and  the painted potential technique based on a time-averaged optical dipole potential \cite{painters}.  Our results are however directly applicable to other systems  such as optical waveguides and photonic lattices \cite{GJ92}, in which curvature-induced potentials \cite{CIP10,Schleich13}, reflectionless potentials \cite{RP10},  and concepts of supersymmetric quantum mechanics \cite{SOW,SUSY14} have already been implemented in the laboratory, and that provide a natural alternative platform to experimentally explore the interplay between geometry and supersymmetry in quantum mechanics.  

\section{Methods}
Let us consider the case in which the superpotential depends on a collective set of parameters $a_0$, $\Phi=\Phi(q_1;a_0)$. The  partner potentials $\V_\pm$ are shape-invariant if they are related by $\V_+(q_1;a_0)=\V_-(q_1;a_1)+R(a_1)$ where the residual term $R(a_1)$ is independent of $q_1$ and $a_1=f(a_0)$ is a new set of parameters obatined form $a_0$ by the action of the function $f(\cdot)$.  By iteration, one can construct the series of Hamiltonians $\{H_k|k=0,1,\dots\}$ with $H_0=H_-$ and $H_1=H_+$, such that $H_s=H_0+\sum_{k=1}^sR(a_k)$, with $a_k=f^k(a_0)$, i.e., obtained by the $f$ function iterated $k$ times.
It follows that the  squared curvatures of the SUSY partner curves $\{\gamma_{s-1},\gamma_{s}\}$ are related by a constant shift
\beqa
\label{SIkappa}
\kappa_{s}^2(q_1)&=&\kappa_{s-1}^2(q_1)-\frac{8m}{\hbar^2}R(a_s)
=\kappa_{0}^2(q_1)-\frac{8m}{\hbar^2}\sum_{k=1}^sR(a_k).\nonumber\\
\eeqa

\vspace{0.21cm}

\noindent
{\bf Acknowledgements:}\\
\noindent
 It is a pleasure to thank X.-W. Guan, E. Passemar, M. Pons, N. Sinitsyn and A. Szameit for stimulating discussions.
This research is supported by the U.S Department of Energy through the LANL/LDRD Program and a  LANL J. Robert Oppenheimer fellowship (AD).

\vspace{0.21cm}

\noindent
{\bf Author Contribution:}\\
\noindent
ADC and AS  initiated the project. ADC developed the theoretical analysis.  MB and ADC
 carried out the numerical simulations. All authors contributed to the analysis and interpretation of the  numerical data and the preparation of the manuscript.

\vspace{0.21cm}

\noindent
{\bf Additional Information}\\
\noindent
{\it Competing Financial Interests:}
The authors declare no competing financial interests.


\begin{thebibliography}{99}




\bibitem{atomchip}  Folman, R. et al.
Controlling cold atoms using nanofabricated surfaces: atom chips. {\it Phys. Rev. Lett.} {\bf 84}, 4749 (2000).

\bibitem{chips}  Reichl, J. $\&$  Vuleti\'c, V. {\it Atom chips} (Wiley-CVH, Singapore, 2011)

\bibitem{ionchip}  Kielpinski, D.,  Monroe, C. $\&$  Wineland, D. J. Architecture for a large-scale ion-trap quantum computer. {\it Nature} {\bf 417}, 709 (2002).

\bibitem{molecularchip}  Meek, S. A.,  Conrad, H. $\&$  Meijer, G. Trapping molecules on a chip. {\it Science} {\bf 324}, 1699 (2009).

\bibitem{electronchip} Hoffrogge, J.,   Fr\"ohlich, R.,  Kasevich, M. A. $\&$  Hommelhoff, P. Microwave guiding of electrons on a chip. {\it Phys. Rev. Lett.} {\bf 106}, 193001 (2011).

\bibitem{painters} Henderson, K., Ryu,  C.,  MacCormick, C. $\&$  Boshier, M. G. Experimental demonstration of painting arbitrary and dynamic potentials for Bose-Einstein condensates. {\it
New J. Phys.} {\bf 11}, 043030 (2009).

\bibitem{Anderson07} Scherer, D. R.,   Weiler, C. N.,  Neely, T. W. $\&$   Anderson, B. P. Vortex formation by merging of multiple trapped Bose-Einstein condensates. {\it  Phys. Rev. Lett.} {\bf 98},
110402 (2007).

\bibitem{Heinzen09}  Liang, J.,   Kohn, R. N.,   Becker, M. F. $\&$  Heinzen, D. 1.5$\%$ root-mean-square flat-intensity laser
beam formed using a binary-amplitude spatial light modulator. {\it Appl. Optics} {\bf 48}, 1955 (2009).

\bibitem{GH12}   Gaunt, A.  L.  $\&$  Hadzibabic, Z. Robust digital holography for ultracold atom trapping. {\it Sci. Rep.} {\bf 2}, 721 (2012).


\bibitem{ring1} Gupta, S., Murch, K. W.,  Moore, K. L.,  Purdy, T. P. $\&$   Stamper-Kurn, D. M. Bose-Einstein condensation in a circular waveguide. {\it Phys. Rev. Lett.} {\bf 95}, 143201 (2005).

\bibitem{ring2}  Olson, S. E.,  Terraciano, M. L.,  Bashkansky, M. $\&$  Fatemi, F. K. Cold-atom confinement in an all-optical dark ring trap. {\it Phys. Rev. A} {\bf 76}, 061404(R) (2007). 

\bibitem{ring3}  Morinaga, M. Circular magneto-optical trap for neutral atoms. {\it J. Phys.  Soc. Japan} {\bf 77}, 104402 (2008). 

\bibitem{ring4}  Heathcote, W. H.,  Nugent, E.,  Sheard, B. T. $\&$   Foot, C. J. A ring trap for ultracold atoms in an RF-dressed state. {\it New J.  Phys.} {\bf 10}, 043012 (2008). 

\bibitem{ring5} Moulder, S.,   Beattie, S., Smith, R. P.,   Tammuz, N. $\&$ Hadzibabic,  Z.  Quantized supercurrent decay in an annular Bose-Einstein condensate. {\it Phys. Rev. A}  {\bf 86}, 013629  (2012).

\bibitem{ring6}  Pritchard, J. D.,  Dinkelaker, A. N.,  Arnold, A. S., Griffin, P. F. $\&$  Riis, E. Demonstration of an inductively coupled ring trap for cold atoms. {\it New J.  Phys.}  {\bf 14}, 103047 (2012).



\bibitem{Wright2013}  Wright, K. C.,  Blakestad, R. B., Lobb, C. J.,  Phillips, W. D. $\&$  Campbell, G. K. Driving phase slips in a superfluid atom circuit with a rotating weak link. {\it Phys. Rev. Lett.} \textbf{110}, 025302 (2013).

\bibitem{Ryu13}  Ryu, C.,  Blackburn, P. W.,  Blinova, A. A. $\&$  Boshier, M. G. Experimental realization of Josephson junctions for an atom SQUID. {\it Phys. Rev. Lett. } {\bf 111}, 205301 (2013).
  
\bibitem{stadium04} Wu, S.,  Rooijakkers, W.,  Striehl, P.  $\&$   Prentiss, M. Bidirectional propagation of cold atoms in a ``stadium''-shaped magnetic guide. {\it Phys. Rev. A} {\bf 70}, 013409 (2004).

\bibitem{Heller06}  Heller, E. J. Guided Gaussian wave packets. {\it Acc. Chem. Res.}  {\bf 39}, 127 (2006).

\bibitem{SRS77}  Switkes, E.,  Russel, E. L. $\&$ Skinner, J. L.  Kinetic energy and path curvature in bound state systems. {\it J. Chem. Phys.} {\bf 67}, 3061 (1977). 

\bibitem{daCosta81}  da Costa,  R. C. T. Quantum mechanics of a constrained particle. {\it Phys. Rev. A} {\bf 23}, 1982 (1981).

\bibitem{daCosta82} da Costa,  R. C. T.  Constraints in quantum mechanics. {\it Phys. Rev. A} {\bf 25}, 2893 (1982).

\bibitem{GJ92}  Goldstone, J. $\&$  Jaffe, R. L. Bound states in twisting tubes. {\it Phys. Rev. B} {\bf 45},  14100 (1992).

\bibitem{CB96} Clark, I. J.  $\&$  Bracken, A. J. Effective potentials of quantum strip waveguides and their dependence upon torsion. {\it J. Phys. A: Math. Gen.} {\bf 29},  339 (1996). 

\bibitem{Clark98} Clark,  I. J. More on effective potentials of quantum strip waveguides. {\it J. Phys. A: Math. Gen.} {\bf 31},  2103 (1998). 

\bibitem{EB89}  Exner, P. $\&$ Seba, P.  Bound states in curved quantum waveguides. {\it J. Math. Phys.} {\bf 30}, 2574 (1989).

\bibitem{EV99}  Exner, P. $\&$  Vugalter, S. A. On the number of particles that a curved quantum waveguide can bind. {\it J. Math. Phys.} {\bf 40}, 4630 (1999).

\bibitem{LP01} Leboeuf, P. $\&$ Pavloff, N.  Bose-Einstein beams: Coherent propagation through a guide. {\it  Phys. Rev. A} {\bf 64}, 033602 (2001).

\bibitem{Schwartz06}  Schwartz, S. et al. 
One-dimensional description of a Bose-Einstein condensate in a rotating closed-loop waveguide.  {\it New J. Phys.} {\bf 8}, 162 (2006).

\bibitem{Struik88} Struik, D. J.  {\it Lectures on classical differential geometry.} (Dover, New York, 1988).

\bibitem{Witten81}  Witten, E. Dynamical breaking of supersymmetry. {\it  Nucl. Phys. B} {\bf 188}, 513 (1981).

\bibitem{CKS95}   Cooper, F.,  Khare, A. $\&$ Sukhatme, U. Supersymmetry and quantum mechanics, 
{\it Phys. Rep.} {\bf 251}, 385 (1995).

\bibitem{DKS88}  Dutt, R., Khare, A. $\&$ Sukhatme, U. P.  Supersymmetry, shape invariance, and exactly solvable potentials, {\it Am. J. Phys.} {\bf 56}, 163 (1988).

\bibitem{Gendenshtein83}  Gendenshte\^in, L. \'E. Derivation of exact spectra of the Schr\"odinger equation by means of supersymmetry. {\it JETP Lett.} {\bf 38}, 356 (1983).
%

\bibitem{propagators}  Pupasov, A. M.,  Samsonov, B. F. $\&$  G\"unther, U. Exact propagators for SUSY partners. {\it  J. Phys. A: Math. Theor.} {\bf 40}, 10557 (2007).

\bibitem{PSU58a} Pokrovskii, V. L., Savvinykh, S. K.  $\&$ Ulinich, F. K.  Super-barrier reflection in the quasiclassical approximation. I. {\it Sov. Phys. JETP} {\bf 34}, 879 (1958).
\bibitem{PSU58b} Pokrovskii, V. L., Savvinykh, S. K.  $\&$ Ulinich, F. K.  Super-barrier reflection in the quasiclassical approximation. II. {\it Sov. Phys. JETP}; {\bf 34}, 1119 (1958).

\bibitem{Henkel96}  Henkel, C.,  Westbrook, C. I. $\&$  Aspect, A. Quantum reflection: atomic matter-wave optics in an attractive exponential potential. {\it J. Opt. Soc. Am. B} {\bf 13}, 233 (1996).

\bibitem{FT04} Friedrich, H.  $\&$ Trost, J.  Working with WKB waves far from the semiclassical limit. {\it Phys.  Rep.} {\bf 397}, 359 (2004).

\bibitem{GAL1}  Guerin, W. et al.
Guided quasicontinuous atom laser. {\it  Phys. Rev. Lett.} {\bf 97}, 200402 (2006).

\bibitem{GAL2} Couvert, A.  et al. 
A quasi-monomode guided atom laser from an all-optical Bose-Einstein condensate. {\it EPL} {\bf 83}, 50001 (2008).

\bibitem{Sukumar86}  Sukumar, C. V. Supersymmetry, potentials with bound states at arbitrary energies and multi-soliton configurations. {\it  J. Phys. A: Math. Gen.} {\bf 19}, 2297 (1986). 

\bibitem{Shabat92}  Shabat, A. The infinite-dimensional dressing dynamical system. {\it  Inverse Prob.} {\bf 8}, 303 (1992).

\bibitem{Spiridonov92}  Spiridonov, V. Exactly solvable potentials and quantum algebras. {\it  Phys. Rev. Lett.} {\bf 69}, 398 (1992).


\bibitem{SK01}  Shevchenko, S. N. $\&$  Kolesnichenko, Yu.  A. Conductance of the elliptically shaped quantum wire. {\it JETP} {\bf 92}, 811 (2001); arXiv:cond-mat/0512110.


\bibitem{Berry96} Berry, M. V. Quantum fractals in boxes.  {\it J. Phys. A} {\bf 26} 6617 1996).

 \bibitem{Talbot1836}  Talbot, H. F. Facts relating to optical science. {\it Philos. Mag.} {\bf 9}, 401 (1836).

\bibitem{BMS01} Berry, M. V., Marzoli, I. $\&$  Schleich, W. P., Quantum carpets, carpets of light. {\it Physics World} (June), 39 (2001); DOI:10.1038/srep02696 (2013)


\bibitem{Schleich1}   Friesch, O. M.,  Marzoli, I. $\&$  Schleich, W. P. Quantum carpets woven by Wigner functions. {\it New J. Phys.} {\bf 2}, 4 (2000).

\bibitem{Schleich2}   Ruostekoski, J.,   Kneer, B.,  Schleich,  W. P.  $\&$  Rempe, G. Interference of a Bose-Einstein condensate in a hard-wall trap: From the nonlinear Talbot effect to the formation of vorticity. {\it Phys. Rev. A} {\bf 63}, 043613 (2001).

\bibitem{Chapman95} 
  Chapman, M. S. et al. 
Near-field imaging of atom diffraction gratings: The atomic Talbot effect.
{\it  Phys. Rev. A} {\bf 51}, R14 (1995).

\bibitem{Mark11}  Mark, M. J. et al. 
Demonstration of the temporal matter-wave Talbot effect for trapped matter waves. {\it  New J. Phys.} {\bf 13}, 085008 (2011). 

\bibitem{Schleich13}  Bittner, S. et al. 
Bound states in sharply bent waveguides: Analytical and experimental approach. 
{\it  Phys. Rev. E} {\bf 87}, 042912 (2013).

\bibitem{CIP10} Szameit, A., et al.
Geometric potential and transport in photonic topological crystals. {\it Phys. Rev. Lett.} {\bf 104}, 150403 (2010).  
\bibitem{RP10} Szameit, A., Dreisow, F., Heinrich, M., Nolte, S. $\&$ Sukhorukov, A. A. Realization of reflectionless potentials in photonic lattices. {\it Phys. Rev. Lett.} {\bf 106}, 193903 (2011).

\bibitem{SOW}  Miri, M.-A.,  Heinrich, M.,  El-Ganainy, R. $\&$  Christodoulides, D. N. Supersymmetric optical structures. {\it Phys. Rev. Lett.} {\bf 110}, 233902 (2013).

\bibitem{SUSY14} Heinrich, M., et al.
Supersymmetric mode converters. {\it Nat. Commun.} {\bf 6}, 3698 (2014).


\end{thebibliography}
\end{document}